# Magnetodielectric coupling and multi-blocking effect in the Ising-chain magnet $Sr_2Ca_2CoMn_2O_9$

T. Basu,[@] N. Sakly, A. Pautrat, F. Veillon, O. Pérez, V. Caignaert, B. Raveau and V. Hardy

Normandie Univ, ENSICAEN, UNICAEN, CNRS, CRISMAT, 14000 Caen, France.

[@]Present Address : School of Physics and Astronomy, University of Minnesota, Minneapolis, MN, 55455, USA

## Abstract

We have demonstrated magnetodielectric (MD) coupling in an Ising-chain magnet $Sr_2Ca_2CoMn_2O_9$, via detailed investigation of ac susceptibility and dielectric constant as a function of temperature, magnetic field and frequency. $Sr_2Ca_2CoMn_2O_9$ consists of spin-chains, made of the regular stacking of one $CoO_6$ trigonal prism with two $MnO_6$ octahedra. The ($Co^{2+}$ $Mn^{4+}$ $Mn^{4+}$) unit stabilizes a (↑↓↑) spin-state along the chains which are distributed on a triangular lattice. This compound undergoes a partially disordered antiferromagnetic transition at $T_N \sim 28$ K. The dielectric constant exhibits a clear peak at $T_N$ only in presence of an external magnetic field ($H \geq 5$ kOe), evidencing the presence of MD coupling, which is further confirmed by field-dependent dielectric measurements. We argue that spatial inversion symmetry can be broken as a result of exchange-striction along each spin chain, inducing uncompensated local dipoles. At low temperatures, a dipolar relaxation phenomenon is observed, bearing strong similarities with the blocking effect typical of the spin dynamics in this compound. Such a spin-dipole relationship is referred to as a 'multi-blocking' effect, in relation with the concept of magnetodielectric "multiglass" previously introduced for related materials.

## I. Introduction

Multiferroic and magnetoelectric materials have been shown to be potential systems for future device applications. Various fascinating magnetoelectric phenomena have been observed in several classes of materials, and different mechanisms have been reported to be responsible for the cross-coupling between spins and dipoles. Initially, it was shown that magnetism-induced ferroelectricity can be obtained in a non-collinear magnetic structure as a result of asymmetric exchange interaction (inverse Dzyaloshinskii-Moriya (D-M) interactions).[1,2] Later on, it was demonstrated that symmetric exchange interaction from collinear magnetic structure could also induce ferroelectricity and magnetoelectric coupling.[3–5] Spin-driven ferroelectricity due to exchange-striction was reported for orthorhombic perovskites $RFeO_3$ (R= rare-earth) with a collinear magnetic structure, on which spin-orbit coupling does not have significant role.[4,6] It is now clear that exchange-striction has a dominant role on spin-dipole coupling and induced ferroelectricity for the multiferroic oxides $RMn_2O_5$ [3,7] and Ising-chain magnet $Ca_3CoMnO_6$.[5], as well as in 1D-organic magnets [8] showing spin-Peierls instability (dimerization due to exchange-striction). Predicted in other systems as well[9], magnetic exchange-striction is also probably responsible for spin-driven pyroelectricity in the non-collinear ferrimagnet $CaBaCo_4O_7$.[10,11]

Among the multiferroic oxides, those containing $Co^{2+}$ are of peculiar interest owing to the potentially very pronounced Ising character of this cation. $Ca_3CoMnO_6$ is the first compound where exchange-striction takes place along with a peculiar ↑↑↓↓ collinear spin structure containing $Mn^{4+}(d^3)$ and high spin $Co^{2+}(d^7)$ ions, which macroscopically generates electrical polarization at the onset of a (short-ranged) magnetic ordering ($T_N \sim 15$ K).[5] This oxide shows a one dimensional Ising character of the spins, with chains of dimeric units built up of $MnO_6$ octahedra and $CoO_6$ trigonal prisms sharing faces. Magnetoelectric coupling and multiferroicity were also observed for the compound $Lu_2CoMnO_6$ with a (↑↑↓↓) spin structure arising from the alternation of $Co^{2+}$ and $Mn^{4+}$ ions located in corner-shared octahedral environments.[12–16]

The spin chain oxides $Sr_{4-x}Ca_xCoMn_2O_9$ (x = 0 to 2.7) [see Ref [17–21]] have an hexagonal structure belonging to the same $A_3MM'O_6$ family as $Ca_3CoMnO_6$, but their chains consist of trimeric units of two $MnO_6$ octahedra sharing faces with one $CoO_6$ trigonal prism (see **Fig. 1**). The title compound $Sr_2Ca_2CoMn_2O_9$ (x=2) exhibits several similarities with $Ca_3CoMnO_6$, since it contains the same spins ($Co^{2+}$ in prisms and $Mn^{4+}$ in octahedra), it also shows Ising-chain magnetism and geometrical frustration, but it strongly differs in its magnetic ground state. Indeed, the different nature of the polyhedral units, "$CoMn_2$" instead of "CoMn", induces different intra-chain interactions. In $Sr_2Ca_2CoMn_2O_9$, it is necessary to take into account 4 exchange couplings $J_i$, leading to a ↑ ($Co^{2+}$) ↓ ($Mn^{4+}$) ↑($Mn^{4+}$) spin structure,[17] compared to $3J_i$ in $Ca_3CoMnO_6$ which shows the (↑↑↓↓) arrangement. Moreover, a long-range ordered (LRO) magnetic state with a clear susceptibility peak at a Néel temperature $T_N$ around 28 K is observed in $Sr_2Ca_2CoMn_2O_9$, unlike $Ca_3CoMnO_6$ which exhibits substantial frequency-dependent susceptibility peak due to domain dynamics, and no peak in the heat-capacity at $T_N$. In other respects, recent studies revealed single-ion magnet (SIM) features in the $Sr_{4-x}Ca_xCoMn_2O_9$ series,[17–20] while these magnetic behaviors are more generally observed in molecular compounds

Therefore, the Ising-chain magnet $Sr_2Ca_2CoMn_2O_9$, exhibits many new interesting features, which tempted us to investigate dielectric, magnetodielectric and possible ferroelectric behavior in this system. Considering the originality of spin dynamics in this material, we have paid special attention for the dynamics of electric dipoles, and compare the spin and dipolar responses over a wide range of frequencies and magnetic fields.

## II. Experimental Details

The sample was prepared by solid-state method as described in Ref.[17] The ac magnetic susceptibility ($h_{ac}$ = 10 Oe) was recorded as a function of temperature at various frequencies (0.01-10 kHz) and in different magnetic fields (0-80 kOe) using an ACMS option of commercial **P**hysical **P**roperties **M**easurements **S**ystem (PPMS, Quantum design). For dielectric measurements, electrodes were painted on the two opposite large faces of the sample (dense platelet of surface (2.0×3.4) mm$^2$ and thickness 0.6 mm) with silver epoxy (Dupont 4929N) cured during one day at room temperature. The measurements were then carried out by impedance analyzer (LCR meter, E4886A, Agilent Technologies) as a function of temperature in a large frequency range of 11.1-120 kHz and magnetic field (0-80 kOe) using a home-made cryogenic insert integrated to PPMS. The data reported in this publication have been recorded in the E//H geometry. Measurements for the E⊥H have given similar results in this polycrystalline material. At the low temperature of measurements, the small value of the loss tangent tanδ (T) ≤ 5.10$^{-2}$ confirms the highly insulating nature of this system, and allows to extract the intrinsic dielectric constant.

### III. Results

*Evidence of magnetodielectric coupling*

The $\chi'$(T) and $\chi''$(T) curves (**Fig. 2a** and **Fig. 2b** respectively) recorded at a fixed frequency of 10 kHz in presence of various magnetic fields (0 – 80 kOe) confirm the previous magnetic study[21] and show the significant impact of an applied magnetic field on the magnetism. In absence of magnetic field, $\chi'$ exhibits a peak at $T_N$, followed by a broad hump around 10 K (~$T_b$) which was ascribed to a "blocking" temperature,[21] and then sharply falls below 10 K, while $\chi''$ exhibits a sharp peak at 8 K and a very weak peak around 28 K. However, one can distinguish an additional feature slightly above $T_N$ on the $\chi'$(T) and $\chi''$(T) curves in zero-field. It was previously shown that it is a pre-transitional regime corresponding to the formation of short-range 1D ferromagnetic-like segments a few K above the long-range spin ordering*[21]*.

Under application of dc magnetic field of 10 kOe, $T_N$ shifts to 26 K, whereas the low temperature feature $T_b$ remains unchanged. In a dc magnetic field of 30 kOe, a broad peak is observed around 20 K in $\chi'(T)$ followed by a change in slope around 10 K, while $\chi''(T)$ clearly exhibits both features in form of a peak ~ 20 K and ~ 10 K (see **Fig. 2a**). For a field of 50 kOe, the peak at $T_N$ shifts further to lower temperature and superimposes with the 10 K feature, as revealed by the broad bump in $\chi'(T)$. One can state that $T_N$ continuously shifts to lower temperature with increasing magnetic field, whereas the peak position of $T_b$ remains nearly the same. The former feature is consistent with antiferromagnetic inter-chain, while the latter is typical of spin blocking phenomena (for energy barriers exceeding a few tens of kelvin).[17,21]

The temperature evolution of the dielectric constant $\varepsilon'$ (T) (**Fig. 2c**) and dielectric loss tangent $tan\delta$ (T) (**Fig. 2d**) of this oxide for nearly the same 10 kHz (actually 11.1 kHz) frequency and the same applied magnetic fields shows that the dielectric and magnetic properties are closely correlated to each other. The specific 11.1 kHz (instead of exact 10 kHz) was chosen to avoid any interference on dielectric behavior from external electrical noise. However, such a small frequency mismatch has negligible difference in dielectric feature (peak position and magnitude) and therefore one can compare to the magnetic results at 10 kHz. In the absence of magnetic field, $\varepsilon'$ slowly increases with decreasing temperature from 50 K, and then sharply drops below ~ 10 K ($T_b$) down to 2 K. The loss part $tan\delta$ is nearly constant with decreasing $T$ and then sharply increases below ~ 10 K with a peak around 7 K. Thus, $\varepsilon'$ and $tan\delta$ trace the feature ~ $T_b$, mimicking the magnetism, but do not show any peak at $T_N$ in absence of magnetic field, unlike $\chi'(T)$. Under a magnetic field of 1 kOe, no peak is detected at $T_N$ as well (not shown here). In contrast, a clear peak is observed at the onset of magnetic ordering $T_N$ for both real and loss parts of dielectric constant in a magnetic field of 5 kOe. It is followed by a cusp with a sharp drop below a temperature $T \sim T_b$. The magnitude of dielectric constant at $T_N$ and $T_b$ becomes stronger with increasing the magnetic field. The maximum of the dielectric constant in the higher temperature region (peak position at $T_N$) shifts to lower temperature with increasing $H$, that is, $\varepsilon'$ and $\tan\delta$ show a peak below ~ 28 K for 5 kOe, which shifts to 26 K for 10 kOe, as observed in magnetism. The peak position at $T_b$ remains nearly the same when increasing the field up to 10 kOe. In a 30 kOe magnetic field, $\varepsilon'(T)$ yields a clear broad peak around 20 K, in accordance with the shift of $T_N$ expected for AFM ordering. The peak position related to $T_N$ is shifted to lower temperature with increasing further the magnetic field and finally overlaps with the feature at $T_b$ under a very high magnetic field (say 80 kOe). A careful analysis of $\tan\delta$ (T) reveals that two features (at $T_N$ and $T_b$) can be clearly distinguished for $H \leq 50$ kOe, whereas, only one peak is observed around 7 K for $H$=80 kOe, in agreement with ac susceptibility. A one-to-one correspondence between the anomalies observed in magnetic and dielectric data is highlighted in **Fig. 3**, displaying the temperature dependence of magnetic susceptibility and dielectric constant at nearly the same frequency (i.e., 10 and 11.1 kHz, respectively) in 10 kOe.

To further support the spin-dipole coupling, we have measured the fractional change of isothermal dielectric constant ($\Delta\varepsilon' = [\varepsilon'(H) - \varepsilon'(0)] / \varepsilon'(0)$) as a function of magnetic field at different temperatures, as shown in **Fig. 4**. A positive MD effect is observed for all the measured temperatures, including temperatures larger than $T_N$ ~ 28 K. It is still significant in magnitude up to 40 K, and almost vanishes at 70 K. Here, it is worth noting that the magnetic susceptibility measurements indicate that the Curie-Weiss regime of a pure paramagnetic state is reached only for $T \gtrsim 50$-$60$ K (see **Fig.5**), a feature consistent with the persistence of short-range magnetic correlation above $T_N$. It can be inferred that such a coupling in this intermediate temperature range is a manifestation of magneto-elastic coupling (magneto-striction effect) related to the presence of short-range magnetic correlations, as observed in many other frustrated systems, including the Ising-chain cobaltate $Ca_3Co_2O_6$.[22,23] In the ordered state $T < T_N$, the sharp change in $\varepsilon'(H)$ around 20 kOe corresponds to the metamagnetic transition, which originates from the field-induced breaking of the partially disordered antiferromagnetic (PDA) order (due to competition between interchain exchange coupling and Zeeman energy). [17] Note that the polycrystalline nature of our samples broadens the metamagnetic transition over a wide field range.

*Spin and dipolar dynamics in presence of dc magnetic field*

To understand the spin and dipolar dynamics and the correlation between them, we have performed frequency dependent ac susceptibility and dielectric measurements in 10 kOe (see **Fig. 6**). The main signature of frequency dependence is observed in the low temperature range (below $T_b$), both for magnetic susceptibility and dielectric constant which exhibit a shallow maximum followed by a marked drop upon cooling. The maximum in $\varepsilon'(T)$ shifts to higher temperature with increasing frequency, revealing dipolar-relaxation of this system. For 10 kOe, the maximum in $\varepsilon'$ shifts from 10 K to 15 K when increasing the frequency from 11.1 to 120 kHz (see **Fig. 6c**). It is to be noted that the maximum in $\chi'$ for 10 kHz exactly matches with the maximum in dielectric for the very close frequency of 11.1 kHz (see **Fig. 3** for a better view). The relaxation time ($\tau$), calculated from imaginary part of ac susceptibility and dielectric constant (i.e. $\chi''(T, f)$ and $\varepsilon''(T, f)$ (where, $\varepsilon'' = \varepsilon' \times tan\delta$)) is plotted in **Fig. 6a** as a function of *1/T*. In the high temperature limit (T ≳ 5 K), the variation has an Arrhenius form $\tau = \tau_0 \exp(\Delta/T)$, where $\tau_0$ is the pre-exponential factor, i.e. attempt time and $\Delta$ is the activation energy. The fitting in the Arrhenius limit gives $\tau_0 = 3\ 10^{-9}$ sec and $\Delta = 47$ K. The fact that the relaxation times of magnetic and dielectric susceptibilities obey to the same law strongly suggests that both the spin and dipolar relaxation might arise from same mechanism.

The spin-chain compounds $Ca_3CoBO_6$, (B= Co, Mn, Rh) also exhibit frequency dependent spin-relaxation and dipolar-relaxation behavior at low *T* below PDA, [5,22–24] and it was a categorized as "multi-glass-like" behavior, similar to "magnetoelectric multiglass" where dipolar relaxation arises due to spin-dipole coupling in a spin-glass system[25–28] In $Sr_2Ca_2CoMn_2O_9$, we have previously shown[17] that the spin relaxation is driven by a blocking process typical of single-ion magnetism (SIM). In this frame, the departure from the linear law in **Fig. 6c** corresponds to the crossover towards a saturation associated to quantum tunneling at the lowest temperatures. Then, considering that the spin and dipolar relaxations seem to originate from the same underlying mechanism, we will refer to this overall phenomenon as a "*multiblocking*" *effect*.

As a next step, we have tried to measure the electrical polarization via pyroelectric measurements, cooling the sample from 40 K down to 5 K using different voltage poling (corresponding to a maximum electric field of 285 kV/m), with and without applied magnetic field. We were unable to detect a significant pyroelectric current and an electrical polarization. At this point, it should be noted that the polycrystalline nature of the sample can strongly affect its polar response. The statistical averaging of the local polarization in grains decreases the macroscopic polarization and the electrical coercivity can increase due to domain walls pinning at grain boundaries and defects. These two extrinsic effects can impede an efficient poling. Another possibility to explain the absence of ferroelectric response is that a polarization is created along each spin chain, but with opposite directions making them to cancel out as a whole.

### IV. Discussion

In $Ca_3CoMnO_6$, Choi *et al.* reported the rise of electrical polarization below 15 K (regarded as a pseudo-$T_N$) but without associated peak in heat capacity and no peak either in dielectric constant (zero-field).[5] For $Sr_2Ca_2CoMn_2O_9$, we observe both a clear magnetic susceptibility peak at zero magnetic field and a magnetic field induced dielectric peak for $H \geq$ 5 kOe. In this compound, the frequency dependent behavior at low temperature below PDA ($T<T_N$) clearly yields a one-to-one correspondence between spin and dipoles, a feature which seems also to be present in $Ca_3CoMnO_6$.[5]

Let us now address the possible origin of dipole creation in $Sr_2Ca_2CoMn_2O_9$. Exchange-striction driven ferroelectricity was initially proposed in $Ca_3CoMnO_6$,[5] due to competing nearest-neighbor ferromagnetic and next-nearest-neighbor antiferromagnetic exchange interactions. The associated ↑↑↓↓ magnetic ordering is responsible for electrical polarization which arises from uncompensated elemental dipoles between cations of different charges ($Co^{2+}$ and $Mn^{4+}$).[5] Later, it was found from DFT calculations that all exchange interactions are antiferromagnetic,[29] but the most stable spin configuration remains the ↑↑↓↓ order, which generates an exchange-striction mechanism. In our title compound, the spins order ↑↓↑ ($Co^{2+}$ - $Mn^{4+}$ - $Mn^{4+}$) results from a compromise between four exchange couplings $J_1$, $J_2$, $J_3$, $J_4$ (see Ref.SM), In the supplemental material, we address in detail the possibility of an exchange-striction mechanism in $Sr_2Ca_2CoMn_2O_9$. Provided that $J_2$ and $J_4$ are not strictly equal to each other, it is shown that a bonus in exchange energy can be gained by introducing a slight move between two consecutives ↑↓↑ units (trimerization process). The mechanism is schematically depicted in **Fig.7**.

**Fig.8** shows the spin states of $Sr_2Ca_2CoMn_2O_9$ for $T < T_N$ under different magnetic fields. **Fig. 8a** represents the PDA state in the zero-field limit. **Fig. 8b** illustrates the "moderate field" regime ($0 < H < 20$ kOe), i.e. below the metamagnetic transition. Two thirds of the chains remain antiparallel to each other, while the incoherent chains are progressively polarized along the direction of the applied magnetic field. This yields unbalance between the up and down ↑↓↑ units along these chains, which makes both $\chi'$ and $\varepsilon'$ to increase as the field is increased. The $T_N$ is still marked by a peak in magnetic $\chi'$ and dielectric susceptibility $\varepsilon'$, whose position shifts to lower values as *H* is increased as typically observed for an AF transition (fig.2). In strictly zero-field, the spin randomness along the incoherent chains must lead to an intrachain compensation of the dipoles associated to the exchange-striction mechanism, a feature which can explain the absence of detectable anomaly in $\varepsilon'$ (T). Applying a dc field allows to partly polarize the spins along these chains, creating a most favorable direction that should be reflected on the dipoles system, inducing a non-zero overall dielectric response. Note that another role of the dc field is to counterbalance (via Zeeman energy) the cost in elastic energy associated to the exchange-striction mechanism, thereby favoring the stabilization of electrical dipoles. **Fig. 8c** shows the "high field" regime, far above the metamagnetic transition, when all the chains are polarized along the same direction. It was previously shown that the metamagnetic transition is expected at ~ 20 kOe when the field is applied along the c-axis.[17] For each grain of a polycrystalline sample, it is the component of the magnetic field along its c axis that must exceed this threshold value. Experimentally, one has observed that the signatures of the $T_N$ start vanishing in applied fields above ~ 50 kOe.[17] The broad maximum that is still present on the $\chi'$(T) and $\varepsilon'$ (T) curves in higher fields (**Fig. 2**) is ascribable only to the blocking effect at $T_b$.

This field driven sequence from PDA/metamagnetic/polarized states can be directly seen on the $\Delta\varepsilon'(H)$ curves at 20 K and 15 K of the **Fig. 4** and is responsible for the S shape and shows consistently the saturation of the MD effect at large field. Interestingly, at the lowest temperatures ($T \leq 10$ K) the blocking effect takes place, and it yields a continuous increase of the MD response as the field is increased, without trend of saturation (at least up to our limit of 80 kOe).

## V. Conclusion

In summary, we have investigated the magnetodielectric properties of the Ising chain-magnet $Sr_2Ca_2CoMn_2O_9$ through frequency dependent ac susceptibility and dielectric constant in presence of different magnetic fields as a function of temperature. A direct coupling between spin dynamics and dipole dynamics is evidenced by the same time scales measured in the dipolar relaxation and spin-relaxation below PDA, which is attributed to "**multiblocking**" effect. We discuss and argue on an exchange-striction origin of the magnetodielectric coupling in this exciting compound.

**Supplementary Material:**

We describe the possible magnetic ground state and origin of exchange-striction which is responsible to create local dipole and possibility of ferroelectric polarization.

**Acknowledgement:**

We would like to thank Laurence Hervé (CRISMAT) for her work on the sample synthesis, as well as the ANR (ANR-16-CE08-0023) and the Région Normandie (Ph D thesis of N. Sakly) for their financial supports.

**Data Availability:**

The data that supports the findings of this study are available within the article [and its supplementary material].

**Figure Captions:**

**Figure 1:** Schematic structure of the stoichiometric spin chain oxide $Sr_2Ca_2CoMn_2O_9$. The chains distributed on a triangular network are built up of face sharing $CoO_6$ trigonal prisms (magenta) and $MnO_6$ octahedra (cyan). Sr and Ca distributed between chains are omitted for the sake of clarity.

**Figure 2:** Real (a) and imaginary (b) part of ac susceptibility as a function of temperature for various fields from 0 to 80 kOe and a fixed frequency of 10 kHz. (c) and (d) are the real part and loss tangent of dielectric constant, for the same fields and for a fixed frequency of 11.1

kHz, respectively. The arrows mark the PDA transition at $T_N$ and the beginning of the blocking process at $T_b$, in the case of zero-field.

**Figure 3:** In-phase (a) and out-of-phase (b) magnetic susceptibility recorded in 1T at a frequency of 10 kHz. In-phase (c) and out-of-phase (b) dielectric constant recorded in the same magnetic field and at a very close frequency (11.1 kHz). The dashed lines show the correspondences between the magnetic and dielectric data.

**Figure 4:** Fractional change of dielectric constant as a function of magnetic field at selected temperatures and a fixed frequency of 71 kHz. The inset is an enlargement of $\Delta\varepsilon'$ vs. $H$ at 15 K and 30 K in order to highlight the change of shape around 20 kOe.

**Figure 5** : Dc-susceptibility curves (*M/H*) recorded in Field Cooled Cooling (FCC) modes, in which the specimen is cooled under application of magnetic field from paramagnetic region (100 K) and data is taken in cooling down to 2K, for a series of magnetic fields. A genuine paramagnetic behavior is recovered only at T~50-60 K, i.e., significantly above the Néel temperature. Dc-susceptibility curves (*M/H*) recorded in the Field Cooled Cooling (FCC) mode, for selected values of the magnetic field. In the FCC mode, the measurements take place upon cooling the sample in the applied magnetic field, from the paramagnetic regime (100 K) down to 2 K. One observes that a genuine paramagnetic behavior (i.e., superimposition of all these curves) is only observed above $T \sim 50\text{-}60$ K, i.e., significantly above the Néel temperature ($T_N \sim 28$ K).

**Figure 6:** The panel (a) reports the relaxation times derived from the maxima of $\chi''(T, f)$ and $\varepsilon''(T, f)$ in both zero-field and 1 T. The red line is a fitting of the high-$T$ part of these data to an Arrhenius law [$\tau = \tau_0 \exp(\Delta/T)$] leading to $\Delta \sim 47$ K and $\tau_0 \sim 3 \cdot 10^{-9}$ sec. The panels (b) and (c) display the in-phase magnetic susceptibility and dielectric constant measured in 1 T at various frequencies. The yellow boxes highlight the *T*-range over which the blocking effect takes place

(starting below a frequency-dependent blocking temperature $T_b$). The peak at higher temperature marks the $T_N$ whose position is virtually frequency-independent.

**Figure 7:** Schematic representation of the alternation of $Co^{2+}$ and $Mn^{4+}$ along the chains. Panel (a) highlights the four types of coupling that must be taken into account. Panel (b) corresponds to the paramagnetic state; the spins are not ordered and the electric dipole between $Co^{2+}$ and $Mn^{4+}$ cancel each other out (white arrows). (c) Spin ordering taking place for two thirds of the chains below $T_N$. (d) Exchange striction accompanying the spin ordering (see text), which induces unbalance between the electric dipoles around each $Co^{2+}$ (green arrows). This yields the appearance of a net polarization along the chain.

**Figure 8:** Top views of the triangular chain lattice, for various regimes of applied field, at $T < T_N$. The symbols + and − correspond to chain obeying ↑↓↑ ordering (as shown in Fig. 7c) with the $Co^{2+}$ spin oriented up or down respectively. The symbols 0 refer to the incoherent chains of the PDA order, having a zero net magnetization. (a) PDA state in zero-field; (b) Moderate magnetic fields ($0 < H < 20$ kOe) tend to polarize the incoherent chains without affecting the antiferromagnetic coupling between + and − chains; (c) High fields ($H \gg 20$ kOe) can break the antiferromagnetic interchain coupling, leading to full polarization of all the chains (in a polycrystalline sample, this regime is reached progressively, depending on the orientation of the $c$-axis of each grain with respect to the direction of the applied field)

**Figure 1:**

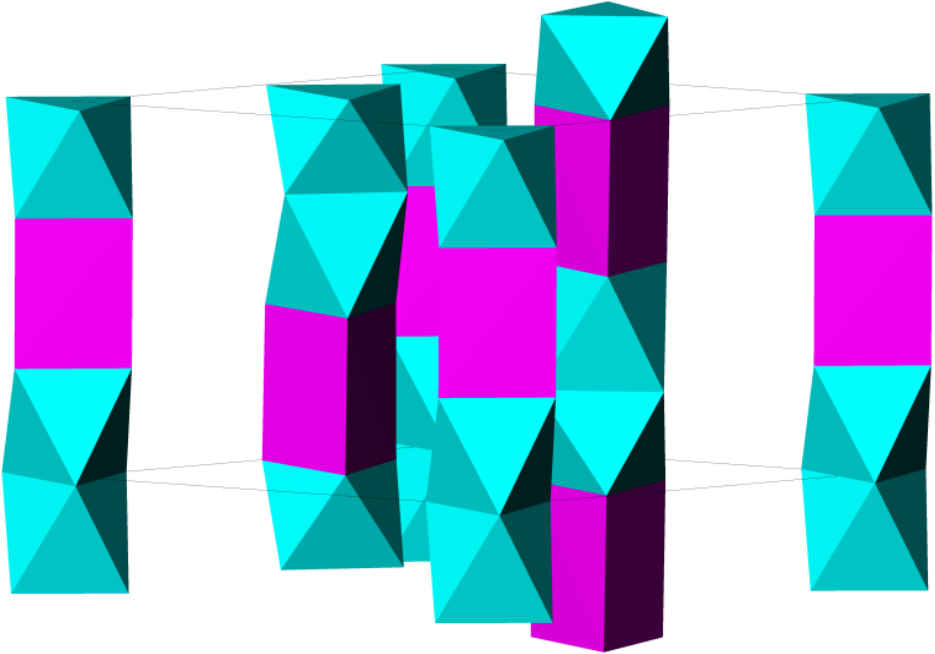

**Figure 2:**

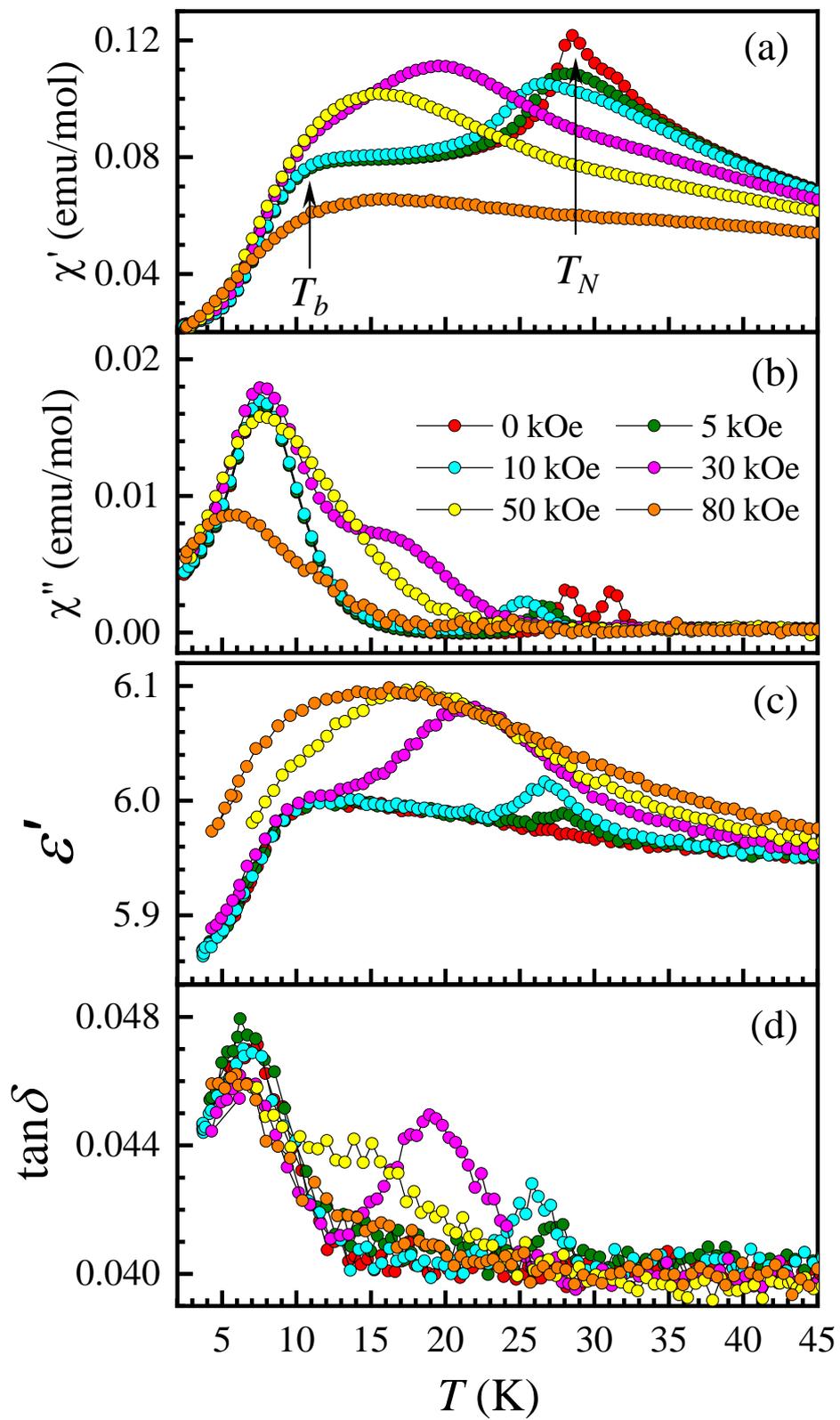

**Figure 3:**

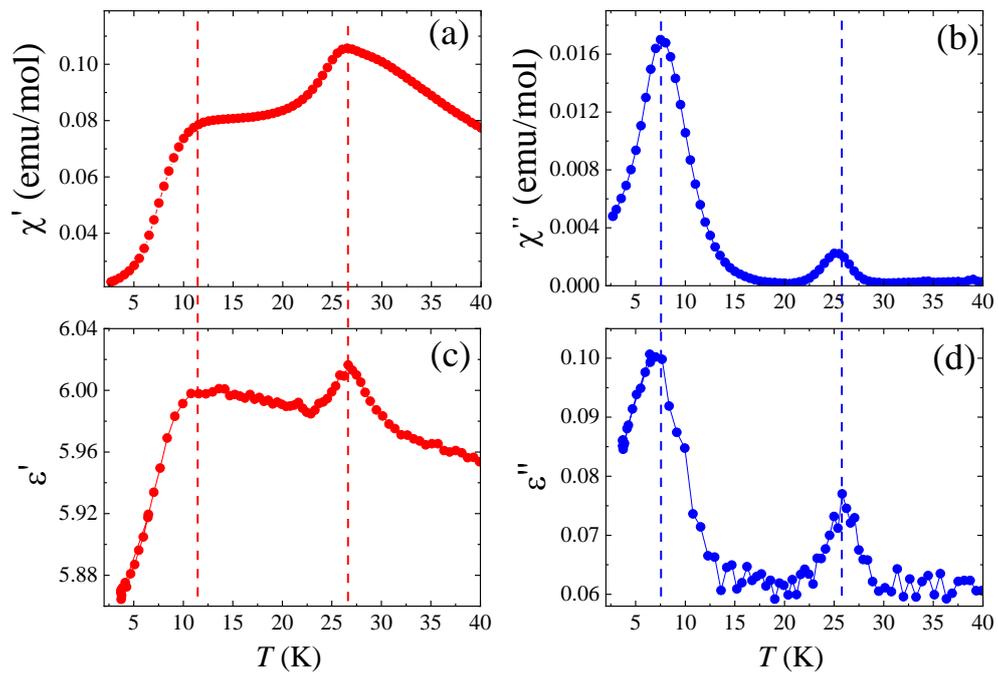

**Figure 4:**

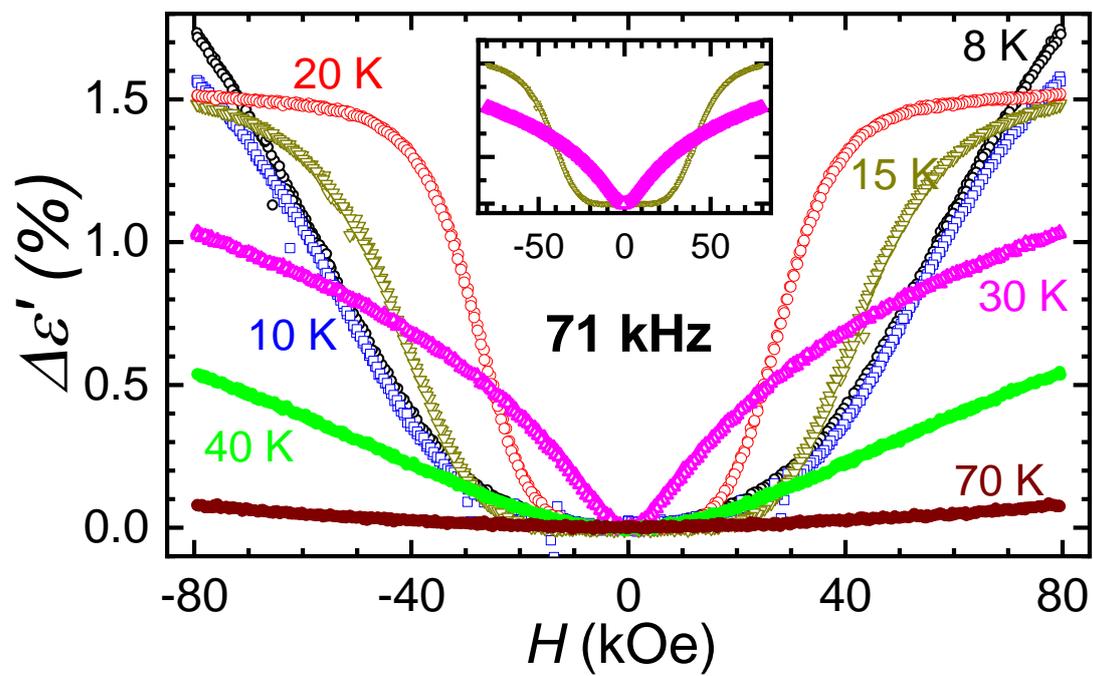

**Figure 5:**

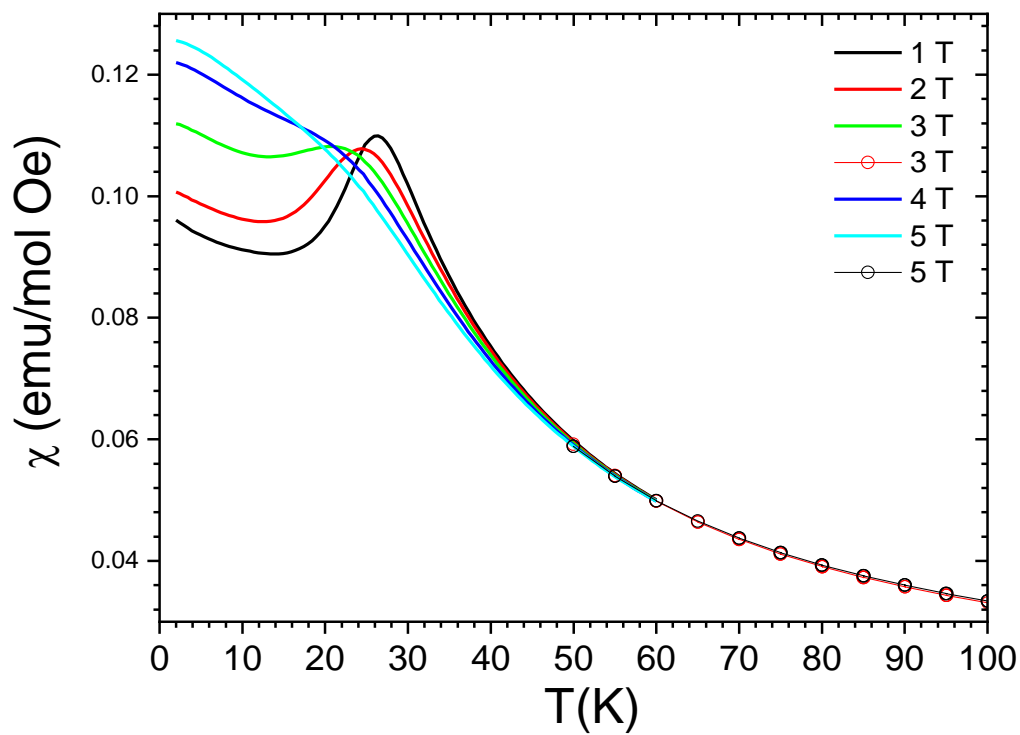

**Figure 6:**

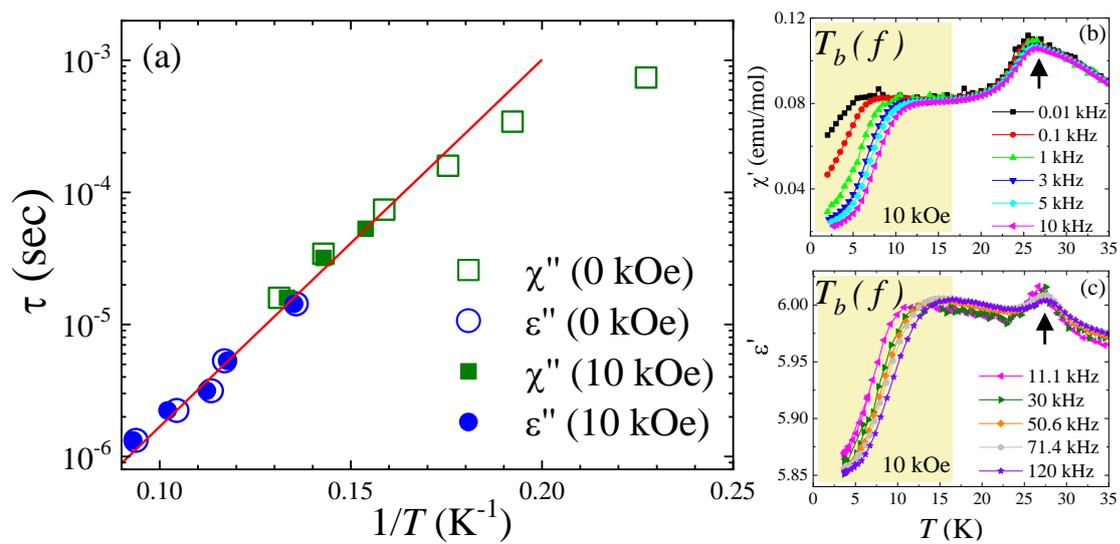

**Figure 7:**

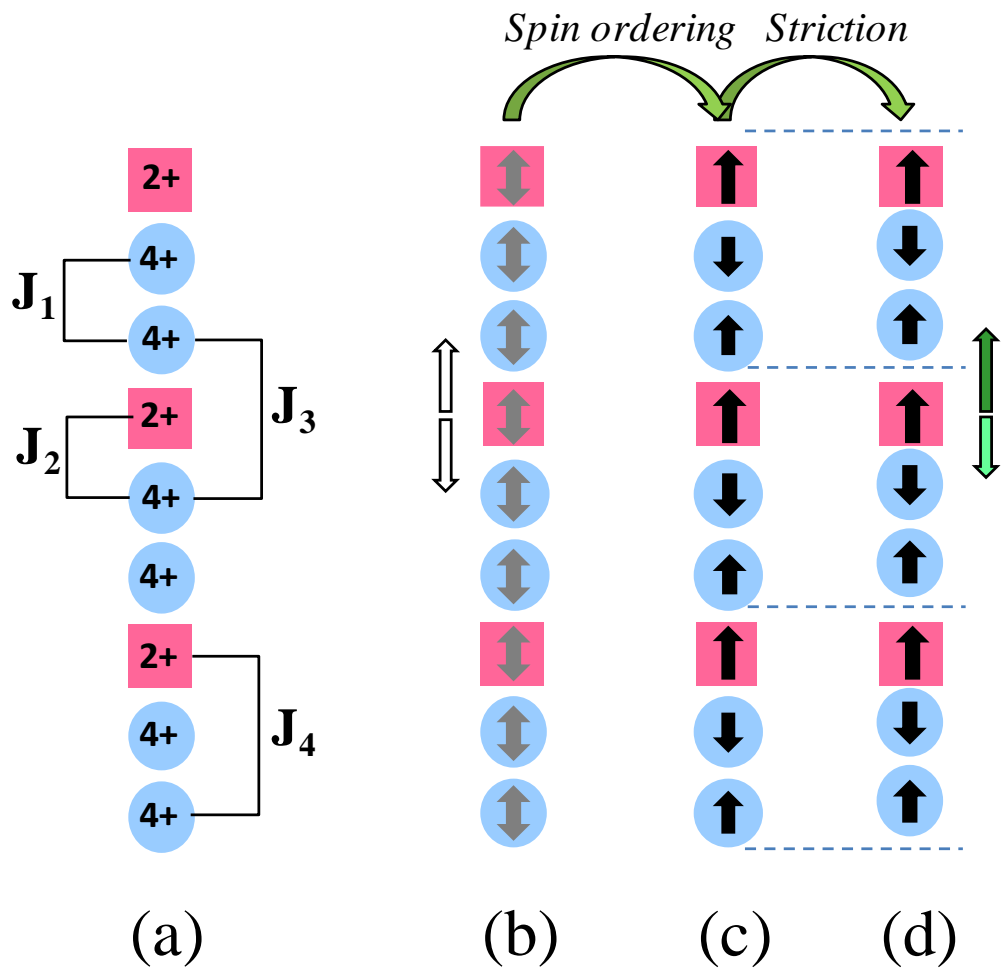

**Figure 8:**

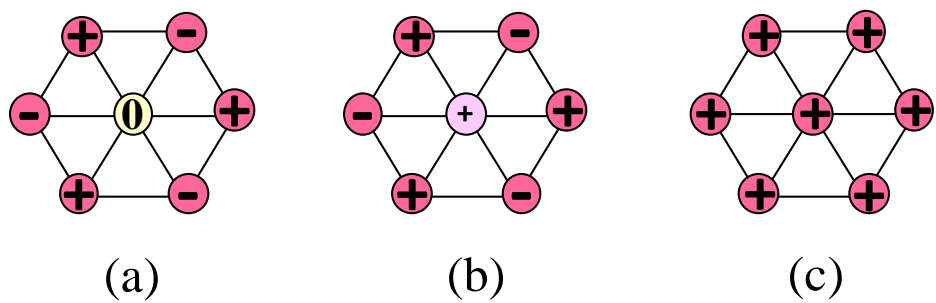

Cluster-Glass Fe2TiO5. *Phys. Rev. B* **2014**, *90* (14), 144426. https://doi.org/10.1103/PhysRevB.90.144426.

(29) Zhang, Y.; Xiang, H. J.; Whangbo, M.-H. Interplay between Jahn-Teller Instability, Uniaxial Magnetism, and Ferroelectricity in Ca3CoMnO6. *Phys. Rev. B* **2009**, *79* (5), 054432. https://doi.org/10.1103/PhysRevB.79.054432.

# Supplemental Material

*ARGUMENTS FOR AN EXCHANGE STRICTION MECHANISM AS THE ORIGIN OF ELECTRICAL DIPOLE EMERGENCE IN $Sr_2Ca_2CoMn_2O_9$*

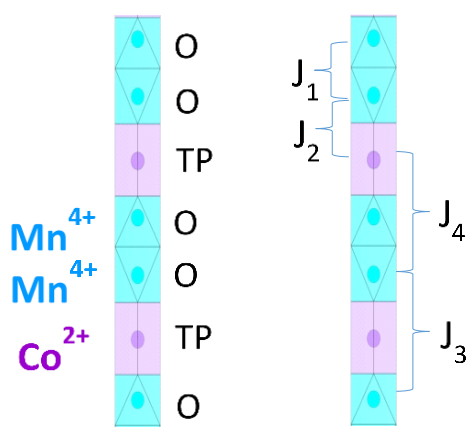

Fig. A1 : Side views of the type of spin chains present in $Sr_2Ca_2CoMn_2O_9$. Left panel emphasizes their structure, that is made of a regular stacking of face-sharing octahedra (noted O, in cyan) hosting $Mn^{4+}$, and of trigonal prisms (noted TP, in magenta) hosting $Co^{2+}$. Right panel indicates the four types of exchange interactions involved in the derivation of the intrachain spin configuration. They corresponds to first or second neighbours, which can also be referred to as nearest-neighbours (nn) or next-nearest-neighbours (nnn), respectively.

Fig A1 displays the structure of the spin chains of $Sr_2Ca_2CoMn_2O_9$, specifying the nature of the polyhedra and of the cations (left), as well as the nature of the first and second neighbours interactions (right).

The third-neighbours interactions within the chains can be considered to be significantly weaker, as attested to by the value of the $Co^{2+}$-$Co^{2+}$ coupling (~ 1 K) determined in the isostructural $Ba_4CoPt_2O_9$ (where $Pt^{4+}$ is nonmagnetic) [N. Sakly, V. Caignaert, O. Perez, L. Herve, B. Raveau, V. Hardy. J. Magn. Magn. Mater. 508, 166877 (2020)]. Previous works on related compounds showed that the coupling $J_1$, $J_2$, and $J_3$ are all antiferromagnetic (AF). Quantitative estimates of the interactions ($J_1$, $J_2$, $J_3$) can be derived from studies on $Sr_4Mn_2NiO_9$ and $Ca_3CoMnO_6$ [A. El Abed, E. Gaudin, J. Darriet, and M.-H. Whangbo, J. Solid State Chem. 163, 513 (2002); Y. Zhang, H. J. Xiang, and M.-H. Whangbo, Phys. Rev. B 79, 054432 (2009).]. Adopting the convention « $2k_B J_i\, S_j S_{j+1}$ » for the exchange energies (with $S = 3/2$ for both $Co^{2+}$ and $Mn^{4+}$) one obtains $J_1 \sim 35$ K, $J_2 \sim 26$ K, and $J_3 \sim 27$ K. No estimate of $J_4$ is available, but this coupling can be expected to be AF (like $J_1$, $J_2$, $J_3$) and of the same order of magnitude.

One is thus facing a situation of geometrical frustration along the chains. In practice, a certain spin configuration should be favored by the competition between the four couplings. Finding this most favorable compromise in terms of interaction energy is the purpose of the following analysis.

**<u>Magnetic ground state</u>**

First of all, let us recall that, owing to the strong Ising character of the $Co^{2+}$ in TP (transmitted to the $Mn^{4+}$ via the magnetic interactions), one can consider only two possible states for each spin, i.e., either ↑ or ↓ along the easy axis (i.e., oriented along the chain direction). Second, since the most favorable spin configuration is driven by the competition between nn and nnn interactions, its determination requires to consider two *nuclear repeat units*. Let us choose Mn-Co-Mn as this basic nuclear unit ; the portion of spin chain used for the calculation of the exchange energy is materialized by the yellow box in Fig. A2 (for practical convenience, the spins direction is drawn perpendicularly to the chain, but this does not affect the calculations).

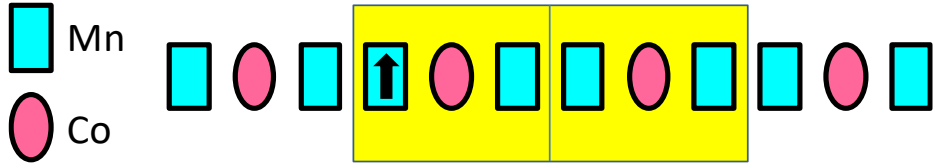

Fig. A2 : Schematic picture of the chain structure, made of a regular stacking between $Mn^{4+}$ (in octahedra) and $Co^{2+}$ (in trigonal prisms). The « spin repeat unit » is highlighted by the yellow box which contains two « nuclear repeat units ». The spin of the first site is arbitrarily chosen to be *up* (↑).

With six adjacent spin sites, there are in principle 64 possible configurations, but one can fix the orientation of one of the spins (↑) without loss of generality ; this reduces the number of configurations to 32. Then, since the interactions are predominantly AF, the configurations of spins with (6 ↑) or (5 ↑ & 1 ↓) are very unlikely. So, we will limit ourselves to the configurations (4↑ & 2 ↓) and (3 ↑ & 3 ↓). Eliminating the equivalent configurations (via horizontal or vertical mirrors) one is left with only 15 different spin configurations. Making use of the estimates of $(J_1, J_2, J_3)$, the exchange energies associated to each of them is plotted versus $J_4$ in Fig. A3.

One observes that the groundstate can correspond to four different spin configurations, depending on the $J_4$ value. An enlargement of Fig. A3 limited to these potential groundstates is shown on Fig. A4, along with the nature of these four spin states. When considering the most likely values of $J_4$ (AF interaction of amplitude close to the other nnn interaction, i.e. $J_3$), it remains two possible configurations which are characterized by the same exchange energy independent of $J_4$. These configurations labelled ① and ③ are shown in Fig. A4.

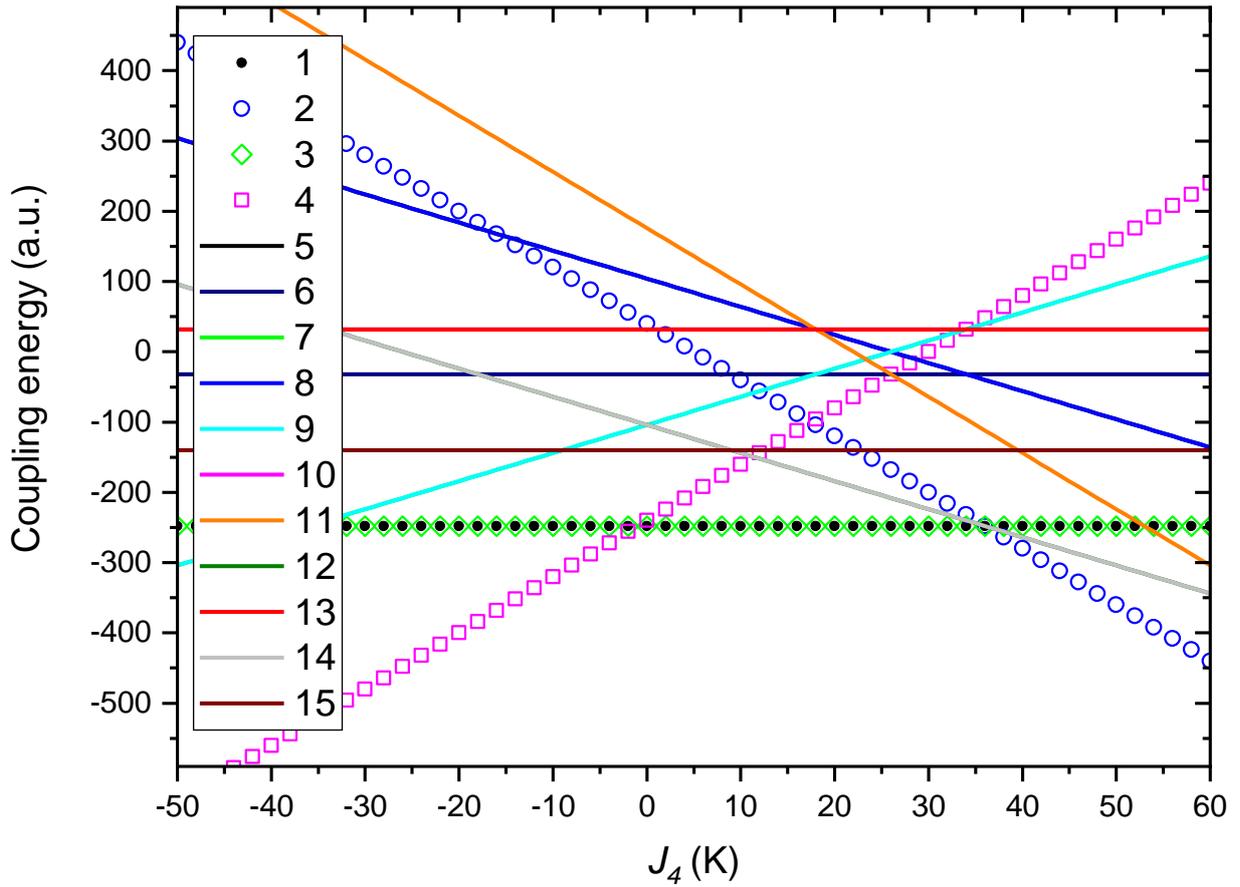

Fig A3 : Exchange energy associated to the spin repeat unit of Fig A2, calculated for the 15 possible spin configurations, as a function of the unknown parameter $J_4$. For ($J_1$, $J_2$, $J_3$), we used the estimates derived from parent compounds (see text).

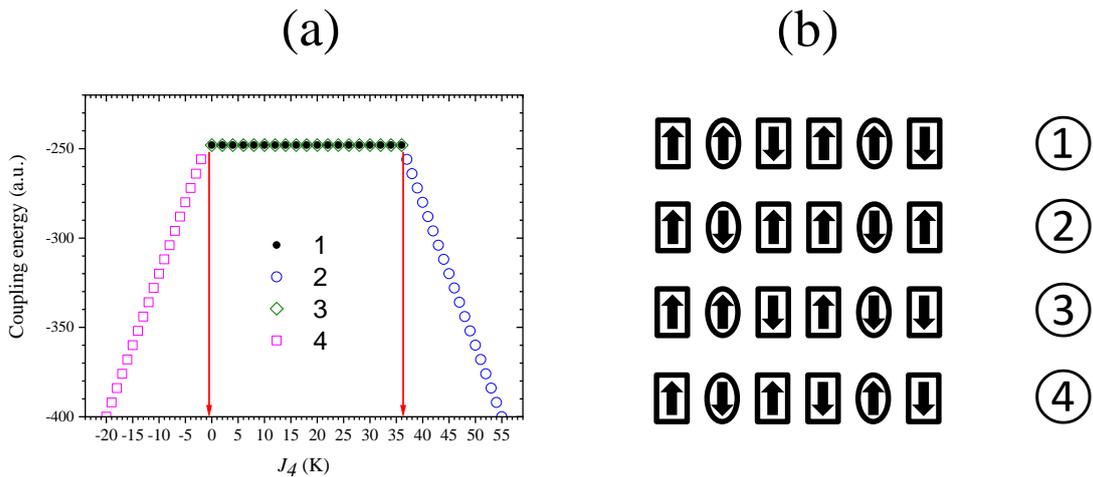

Fig A4 : (a) Enlargement of Fig. A3 focusing on the four configurations susceptible to be the groundstate. These spin configurations are displayed in panel (b) (ellipses are $Co^{2+}$ and rectangles are $Mn^{4+}$).

The spin configuration ① actually corresponds to the (↑↓↑) order previously reported in paper [1] (referring to a Co-Mn-Mn unit). It has a net magnetization associated to the $Co^{2+}$ spins. The spin configuration ③ corresponds to an AF order ; Such a magnetic structure should be characterized by a doubling of parameter (not found experimentally in neutron diffraction experiments) and it should have a zero net magnetization (at odds with the experimental magnetization data). Accordingly, the only spin configuration that is consistent with the present analysis and the experimental data is the configuration ①. Even though ① and ③ are degenerate in zero field (and when limiting the interactions up to second neighbours), it can be noted that ① is favored by application of magnetic field, owing to the Zeeman energy term.

**Exchange-striction phenomenon**

Let us now address the origin of a magnetodielectric response in $Sr_2Ca_2CoMn_2O_9$. In what follows, we investigate the possibility of a mechanism similar to that described in the parent compound $Ca_3CoMnO_6$, where electrical polarization originates from uncompensated elemental dipoles between cations of different charges ($Co^{2+}$ and $Mn^{4+}$). The underlying mechanism is basically an exchange-striction phenomenon in which the spin system can lower its exchange energy by moving the positions between some of the cations along the chains.

Looking at the spin configuration ①, one observes that both the interactions $J_1$ and $J_3$ are fully satisfied, while it is not the case for $J_2$ and $J_4$. Thereafter, we limit ourselves to the mechanisms susceptible to induce electrical dipoles, i.e. various bound lengths between $Co^{2+}$ and $Mn^{4+}$. We also make use of the reasonable assumption that the coupling intensity between any pair of spins increases as the interdistance between them is decreased. Starting from the spin configuration ①, it turns out that the exchange energy can be further optimized (i.e. decreased) by moving apart the spin at the tick positions shown in Fig. A5, either at the upper ones (full lines) to gain energy via $J_2$ or at the lower ones (dashed lines) to gain energy via $J_4$. To be more quantitative, let us adopt the following assumptions : (i) The overall chain length is kept constant and the Co positions are fixed ; (ii) There is an overall shift of the $Mn_2$ pair; one Mn moves away from one neighboring Co ,whereas the second Mn becomes closer to the second neighboring Co ; (iii) Accordingly, $J_1$ and $J_3$ are not modified, while $J_2$ splits into ($J_2 - dJ_2$) and ($J_2 + dJ_2$), as well as $J_4$ into ($J_4 - dJ_4$) and ($J_4 + dJ_4$) ; (iv) All the $J_i$ and the $dJ_i$ are positive.

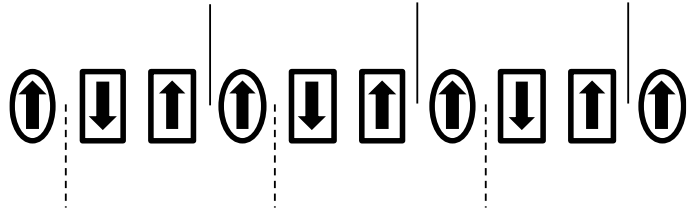

Fig. A5 : Groundstate spin configuration ① with $Co^{2+}$ in ellipses and $Mn^{4+}$ in rectangles The ticks mark the positions for which extra spacings can be beneficial in terms of exchange energy.

Calculations of the exchange energies are then carried out as done previously. With increased intercationic separation at the upper ticks, the exchange energy turns to $E_{ex} = E_①- 8(dJ_2 - dJ_4)$. Such a distortion is thus authorized if $dJ_2 > dJ_4$, leading to the modified spin structure (a) in Fig. A6. For spacings at the lower ticks, $E_{ex} = E_①- 8(dJ_4 - dJ_2)$, making such a distortion favorable for $dJ_4 > dJ_2$. This induces the structure (b) in Fig . A6, which also corresponds to (b') (equivalent via a vertical mirror).

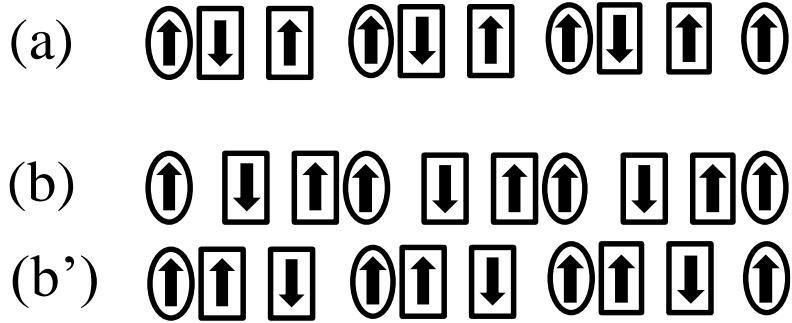

Fig. A6 : Schematic pictures of the spin chain structures resulting from exchange-striction effects applied to the groundstate configuration ① (see text).

In both cases, one observes the same kind of trimerization involving Co-Mn-Mn units. In practice, one can thus reasonably expect that a exchange-striction mechanism [either (a) or (b)] can take place in $Sr_2Ca_2CoMn_2O_9$ provided that $J_2$ is not strictly equal to $J_4$. At this stage, it must be specified that we have only dealt with the exchange energy in the above analysis. In principle, the onset of magnetostriction also requires that the *gain in magnetic energy exceeds the cost in elastic energy* (associated to the distortion of the chain structure). This second energy term is not taken into account in the present approach.